\documentclass[12pt,preprint]{aastex}
\usepackage{psfig} 

\def\be{\begin{equation}}
\def\ee{\end{equation}}

\def\m{~$\mu$m}

\def\HII{\ion{H}{2}}

\def\CII{[\ion{C}{2}]}
\def\OI{[\ion{O}{1}]}

\def\Htwo{H$_2$}

\def\ISO{{\it ISO}}
\def\IRAS{{\it IRAS}}

\def\IRAScolor{${f_\nu (60 \mu {\rm m})} \over {f_\nu (100 \mu {\rm m})}$}

\begin {document}

\title{\OI~63\m\ Emission from High and Low Luminosity AGN Galaxies}
 
\author{Daniel A. Dale,\altaffilmark{1,2} George Helou,\altaffilmark{2,3} James R. Brauher,\altaffilmark{3} Roc M. Cutri,\altaffilmark{2} Sangeeta Malhotra,\altaffilmark{4} and Charles A. Beichman\altaffilmark{5}}
\altaffiltext{1}{\scriptsize Department of Physics and Astronomy, University of Wyoming, Laramie, WY 82071}
\altaffiltext{2}{\scriptsize Infrared Processing and Analysis Center, California Institute of Technology, M.S. 100-22, Pasadena, CA 91125}
\altaffiltext{3}{\scriptsize SIRTF Science Center, California Institute of Technology, M.S. 220-6, Pasadena, CA 91125}
\altaffiltext{4}{\scriptsize Space Telescope Science Institute, 3700 San Martin Drive, Baltimore, MD 21218}
\altaffiltext{5}{\scriptsize Jet Propulsion Laboratory, California Institute of Technology, 4800 Oak Grove Drive, Pasadena, CA 91109}

\begin {abstract}
The {\it Infrared Space Observatory} was used to search for a tracer of the warm and dense neutral interstellar medium, the \OI~63.18\m\ line, in four ultraluminous \IRAS\ sources lying at redshifts between 0.6 and 1.4.  While these sources are quasars, their infrared continuum emission suggests a substantial interstellar medium.  No \OI\ flux was securely detected after probing down to a 3$\sigma$ sensitivity level sufficient for detecting line emission in starbursts with similar continuum emission.  However, if the detection threshold is slightly relaxed, one target is detected with 2.7$\sigma$ significance.  For this radio-quiet quasar there is likely a substantial dense and warm interstellar medium; the upper limits for the three radio-loud sources do not preclude the same conclusion.  Using a new, uniformly-processed database of the $ISO$ extragalactic far-infrared spectroscopy observations, it is shown that nearby Seyfert galaxies typically have higher \OI-to-far-infrared ratios than do normal star-forming galaxies, so the lack of strong \OI~63\m\ emission from these high-redshift ultraluminous sources cannot be attributed to their active cores.
\end {abstract}
 

\section {INTRODUCTION}

Ultraluminous infrared sources were brought into prominence by the {\it Infrared Astronomical Satellite}, or \IRAS.  Nearby Arp~220 is the prototypical ultraluminous infrared galaxy (Soifer et al. 1984).  The nature of the main power source in such objects is still subject to debate, with some favoring dust-enshrouded active galactic nuclei and others supporting optically-thick starbursts.  Though the point is still debated, recent evidence points towards star formation being the main engine (Genzel et al. 1998; Rowan-Robinson 2000; Surace \& Sanders 2000; Colina et al. 2001; Tran 2001; Farrah et al. 2003), suggesting that starburst properties may apply broadly to infrared-luminous objects.

A small subset of \IRAS\ objects detected at around 1~Jy at 25 and 60\m\ were subsequently shown to be at high redshifts.  Some of them became known as hyperluminous infrared galaxies because the infrared emission dominates the bolometric power radiated by these sources and exceeds 10$^{13}~L_\odot$ (e.g. Sanders \& Mirabel 1996).  Although most of these objects are classified as quasars and BL~Lacertae objects, their large ratios of infrared-to-optical luminosity are akin to ratios seen in nearby ultraluminous infrared galaxies.  In addition, recent surveys of low-redshift quasi-stellar objects have detected strong CO emission in a majority of the host galaxies (e.g. Evans et al. 2001; Scoville et al. 2003).  Interestingly, the survey led by Scoville does not focus on infrared-excess sources, indicating that dense interstellar media may be commonplace for quasars.  These pieces of evidence suggest that a large fraction of the bolometric luminosity in high-redshift ultraluminous infrared galaxies may arise from star formation, making them the elusive missing link between pure star-forming sources and true quasars.  Their physics are therefore of particular interest to elucidate.


Deciphering the physical processes within those sources at high redshift is complicated by heavy dust obscuration, making dust-penetrating far-infrared signatures from the interstellar medium the most natural tools.  As outlined in Hollenbach \& Tielens (1999), the heating of neutral interstellar gas is linked to star formation through the photo-ejection of electrons from polycyclic aromatic hydrocarbons and small dust grains.  The dominant coolers of the neutral interstellar medium are the \CII~158\m\ and \OI~63\m\ lines.
A substantial portion of the \CII~158\m\ and \OI~63\m\ fluxes from galaxies is thought to originate in photo-dissociation regions at the interfaces between \HII\ regions and molecular clouds (see Malhotra et al. 2001 and Contursi et al. 2002 and references therein).  Additional contributions to these lines come from the diffuse neutral medium, and from the diffuse ionized medium and \HII\ regions for \CII~158\m.  Strong \OI\ emission can also be associated with shocks (e.g. Spinoglio \& Malkan 1992; van der Werf et al. 1993).  Infrared fine structure line studies in the Milky Way have traditionally relied on photodissociation region modeling to deduce physical parameters like gas density $n_{\rm e}$, temperature $T$, and  incident far-ultraviolet radiation flux $G_0$.  The cold neutral medium is predominantly cooled by \CII~158\m\ since the ion is abundant and the 158\m\ line is easy to excite ($\Delta E/k \approx 92$~K).  In more dense and warm environments, where 
$G_0 \gtrsim 10^3$ and $n_{\rm e} \gtrsim 10^5~{\rm cm}^{-3}$, \OI~63\m\ becomes the dominant coolant of the interstellar gas; \OI~63\m\ lies $\Delta E/k \approx 228$~K above the ground state (Hollenbach \& Tielens 1999).  

In the pre-{\it Infrared Space Observatory} (\ISO) era, {\it Kuiper Airborne Observatory} data already pointed to \OI~63\m\ as a major cooling line of the warm, neutral, dense interstellar medium (Lord et al. 1995).
\ISO\ made it possible to study \OI~63\m\ in ``normal'' galaxies whereas the {\it Kuiper Airborne Observatory} was only used to study starburst galaxies since the atmosphere and warm telescope limited observing to high surface brightness objects.  
In starbursts, \OI~63\m\ accounts for 0.3\% of the total far-infrared output, barely a factor of two below that from \CII~158\m.  \ISO\ data have confirmed this result and shown even higher \OI/$FIR$ ratios in the more ``normal'' galaxies driven by star formation (Malhotra et al. 2001).  Moreover, since \OI~63\m\ requires more energy to excite than \CII~158\m, Malhotra and collaborators have shown that \OI/\CII\ increases with \IRAScolor\ until \OI~63\m\ exceeds \CII~158\m\ in luminosity for galaxies with the warmest interstellar media.  The \OI~63\m\ far-infrared tracer thus offers a unique opportunity to search for the warm and dense interstellar medium in ultraluminous infrared galaxies at high redshifts, many of which are intensely star-forming sources.
In this work \ISO\ observations of four ultraluminous infrared sources in the redshift range $0.59<z<1.40$ are described.  The results are compared to what is observed for local Seyferts and star-forming galaxies.

\section {THE SAMPLE}
\label{sec:sample}

The sources were selected according to redshift and \IRAS\ flux.  A sky-wide search of the NED database yields about 40 
\IRAS\ sources optically identified and measured to have redshifts $0.58<z<1.85$, a range which observationally places the rest-frame \OI~63\m\ line between 100 and 180\m.  This redshift selection was driven by the sensitivity spectrum of \ISO's {\it Long Wavelength Spectrometer} (LWS; Clegg et al. 1996; Kessler et al. 1996).
Further examination of each source was carried out according to the raw \IRAS\ data, the Digitized Sky Survey, and the literature.  The subset with robust optical identifications and reliable infrared detections, with $f_\nu(25\micron)>120$~mJy and $f_\nu(60\micron)>250$~mJy,
results in a candidate list of seven good sources.   Four of these sources were ultimately observed for \OI~63\m\ during the \ISO\ mission (see Table~\ref{tab:sample}).

\begin{figure}[!ht]
\plotone{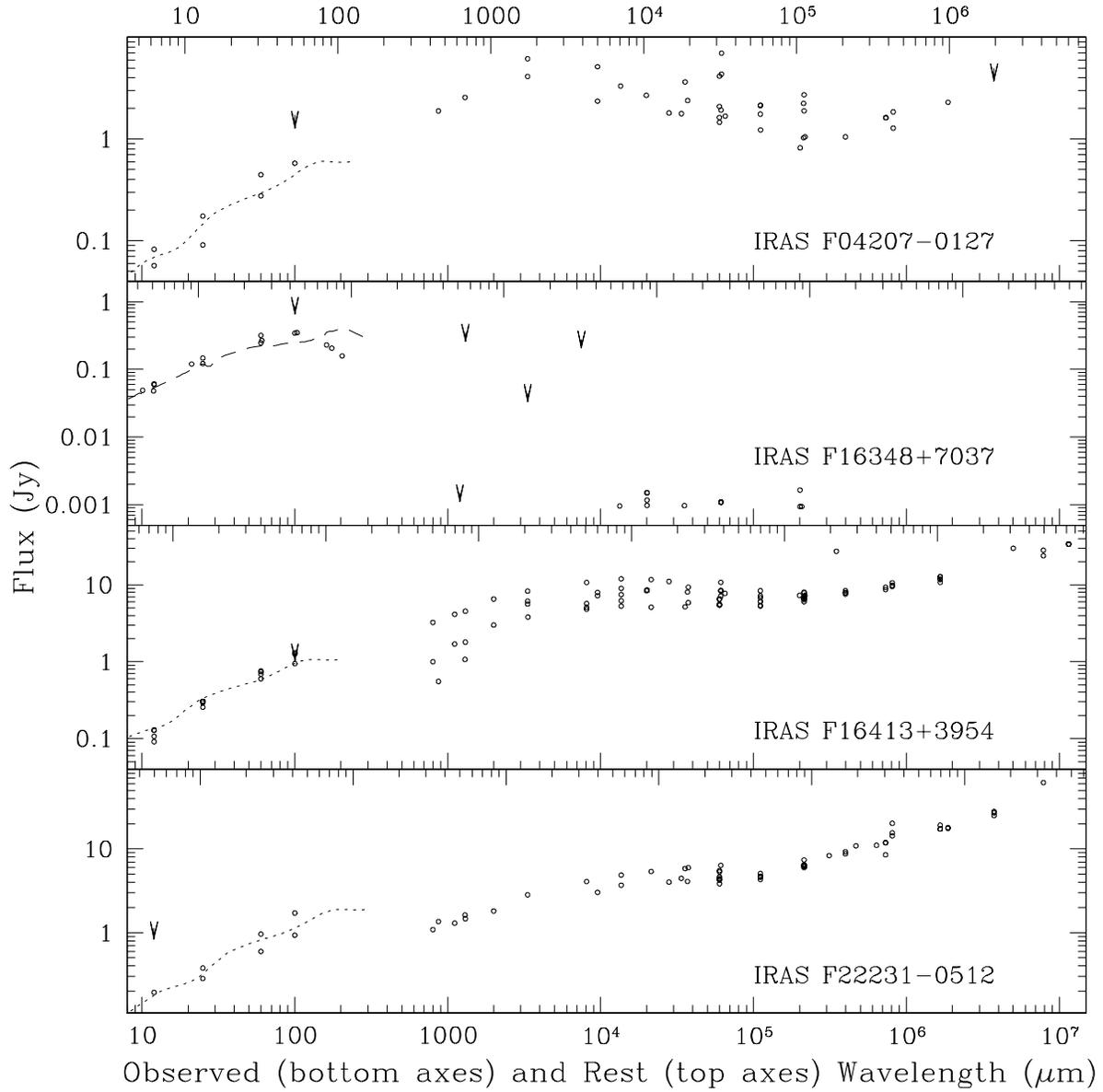}
\caption[] {\ The infrared-radio spectra from NED.  The dotted (dashed) line indicates the average radio-loud (radio-quiet) QSO profile fit to the infrared for the purposes of estimating rest frame 42-122\m\ fluxes (see text).} 
\label{fig:ir_radio_spectra}
\end{figure}

Three of these sources are ``blazars,'' FR~II radio sources with the jet pointed at us, essentially the more luminous quasar equivalent of Bl~Lacs.  All four sources have (rest-frame) far-infrared luminosities that fit in the ultra- or even hyper-luminous infrared category ($>$$10^{12}~L_\odot$ and $>$$10^{13}~L_\odot$).  To estimate the rest-frame far-infrared luminosity for these higher redshift sources, average QSO curves (Elvis et al. 1994) are first normalized to the global mid- and far-infrared fluxes (see Figure~\ref{fig:ir_radio_spectra}) and then the luminosity is integrated from 42 to 122\m\ (according to the classic definition, the far-infrared flux $FIR=1.26\cdot10^{-14}~[2.58 f_\nu(60\micron)+f_\nu(100\micron)]~{\rm W~m^{-2}}$ recovers the integrated 42-122\m\ rest-frame continuum flux for local galaxies; Helou et al. 1988).  

Comparisons with local sources can be problematic due to the redshifting of the spectral energy distributions.  A comparison of the far-infrared to near-infrared continuum luminosity ratio can be made if the observed optical magnitudes for the four $z\sim1$ sources are approximated as their rest-frame $J$ magnitudes.  The sources in this sample have $FIR$-to-$J$ rest-frame luminosity ratios between 0.6 and 23, values that fall within the envelope of normal galaxies, $\sim$0.1 to 25.  The latter range is derived from \IRAS\ and 2MASS data for the \ISO\ Key Project on Normal Galaxies sample (Helou et al. 1996; Dale et al. 2000).

\subsection{IRAS~F04207-0127}

Also known as PKS~0420-012, this radio-loud, flat-spectrum quasar is a core-dominated radio source and likely has beamed synchrotron close to the line-of-sight (Wehrle et al. 1992; Brotherton 1996).  The source is highly variable and polarized in the optical (Webb et al. 1988; Wills et al. 1992), and exhibits superluminal motions at radio wavelengths (Hong et al. 1999; Britzen et al. 2000; Jorstad et al. 2001).  Douglas et al. (1992) did not detect any CO(2$\rightarrow$3) absorption and provide an upper limit to the CO column density as a few times 10$^{15}~{\rm cm}^{-2}$.  Assuming the local CO-to-\Htwo\ ratio of $\sim$$10^{-4}$ applies (Dickman 1978), this upper limit is significantly smaller than the range $10^{16}-10^{18}~{\rm cm}^{-2}$ seen for local star-forming and ultraluminous infrared galaxies (e.g. Young et al. 1995; Solomon et al. 1997; Helfer et al. 2003).


\subsection{IRAS~F16348+7037}

This radio-quiet quasar (PG~1634+706) has an infrared spectral profile that matches the extended AGN dust torus model (Haas et al. 1998).  A search for CO(5$\rightarrow$4) and CO(2$\rightarrow$1) emission did not detect any for this source (Evans et al. 1998; Barvainis et al. 1998), consistent with the lack of support for a starburst component in the galaxy's spectral energy distribution (Rowan-Robinson 2000).  The lower radio emission allows the distinct infrared hump to emerge as a signature of dust emission, much clearer here than for the other three objects in this study.

%

\subsection{IRAS~F16413+3954}

Radio-loud quasar 3C~345 is a core-dominated, flat spectrum radio source and likely has beamed synchrotron (Brotherton 1996).  It is a superluminal quasar variable in the optical, radio, and X-ray (e.g. Unwin et al. 1997).  The bulk of the bolometric luminosity arises from the optical-submillimeter wavelength range (Rantakyr\"o et al. 1998 and references therein).

\subsection{IRAS~F22231-0512}

Like IRAS~F04207-0127 and IRAS~F16413+3954 in this sample, IRAS~F22231-0512 (3C~446) is a radio-loud quasar with a core-dominated radio source and likely has beamed synchrotron (Wills et al. 1995).  Similarly, its optical emission is both variable and polarized (Mead et al. 1990; Barbieri et al. 1990; Hufnagel \& Bregman 1992).  CO(1$\rightarrow$2) was not observed in absorption; the total CO column density 2$\sigma$ upper limit is $3\cdot10^{14}~{\rm cm}^{-2}$ (Drinkwater, Combes, \& Wiklind 1996).

\section{OBSERVATIONS AND DATA REDUCTION}

For each source the LWS02 AOT was used to measure a spectral segment comprising the \OI~63\m\ line at its expected redshift.  The observations for this program were designed to reach sensitivities comparable to the expected \OI~63\m\ emission from star-forming galaxies;
the line strength was predicted according to the infrared continuum flux and typical line-to-continuum ratios for star-forming galaxies (e.g. Malhotra et al. 2001).  This assumption guided sensitivity and integration estimates.  On-source integrations ranged from 2.0 to 5.5 hours.  Precautionary off-source observations that typically lasted one-third of an hour were also taken for each source.  The off-source data ultimately were not used since no strong Milky Way foreground emission or strong continuum gradients needed to be characterized and removed.  Table~\ref{tab:observations} lists the \ISO\ observation identification numbers (TDTs) and relevant details.



Data reduction on the Pipeline~7 datasets was performed using Version~8.1 of the $LWS$ Interactive Analysis software package (Hutchinson et al. 2001).  Post-pipeline processing was facilitated with Version~2.1 of the \ISO\ Spectral Analysis Package (Sturm et al. 1998).  
On-source integrations from the separate TDTs are averaged after a first-order baseline is initially removed.  Full details of the data reduction process is outlined in Brauher (2004).

\section{RESULTS}

\begin{figure}[!ht]
\plotone{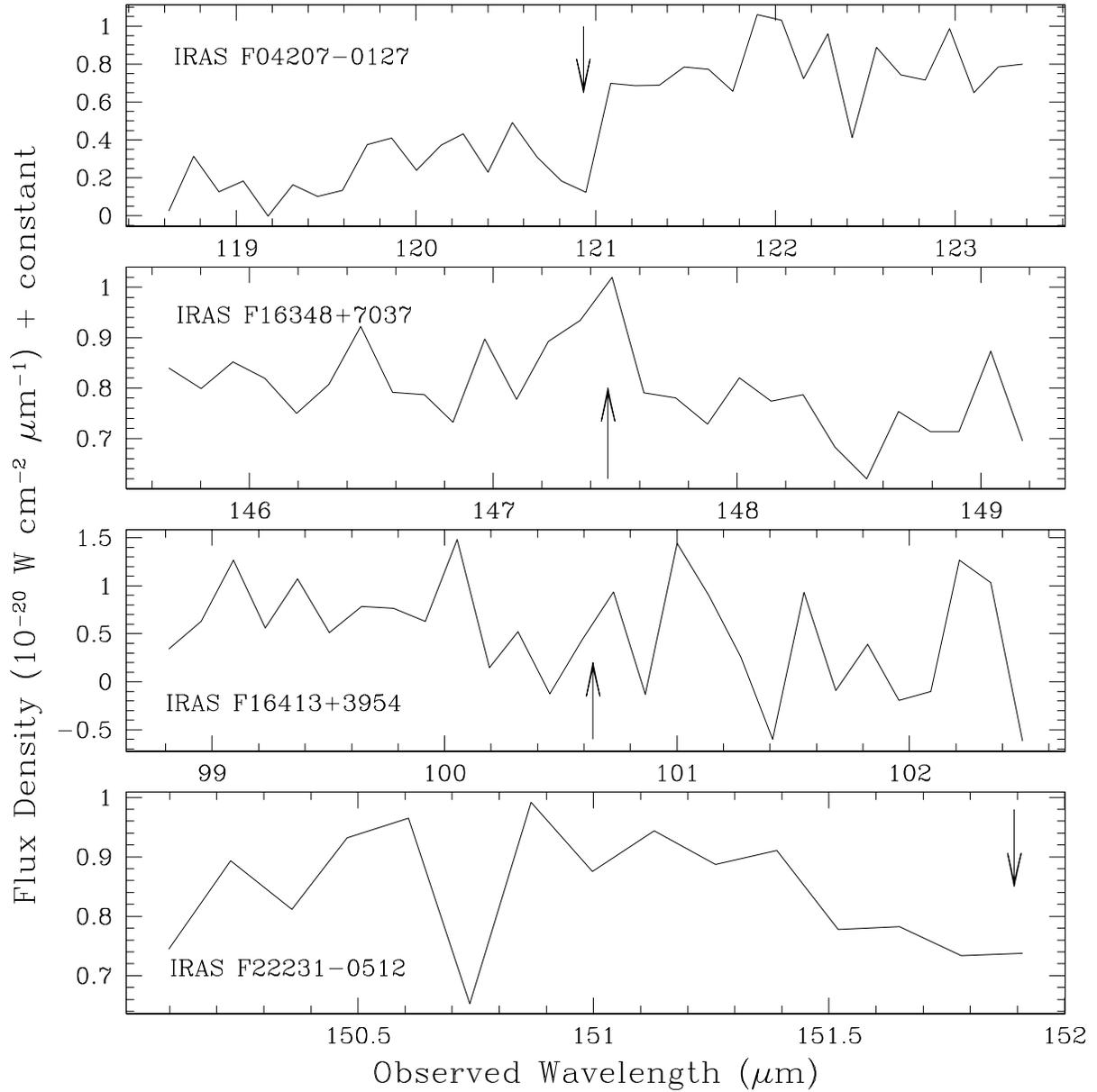}
\caption[] {\ The $ISOLWS$ spectra with the rest-frame \OI~63\m\ wavelength indicated by arrows.} 
\label{fig:lws}
\end{figure}

No robust detections of the \OI~63\m\ line were obtained in the survey (Figure~\ref{fig:lws}).  The redshift of IRASF~22231$-$0512 unfortunately placed the \OI~63\m\ line at the useful edge of the LW3 detector, but the wavelength coverage appears to be sufficient to rule out a detection.  The r.m.s. noise level in each averaged continuum is calculated from the dispersion about a line fitted to the averaged data.  Upper limits to the \OI~63\m\ flux are derived from the r.m.s. values assuming a Gaussian beam of fixed instrumental width (0.60\m\ for the detectors used).  Table~\ref{tab:observations} lists both the r.m.s. and the 3$\sigma$ upper limits.  

For one source, IRAS~F16348$+$7037, there may be a weak detection of \OI~63\m.  A line flux for this source is computed via a composite fit of a first-order baseline and a Gaussian beam of width 0.60\m.  The fitted central wavelength is 147.396\m, corresponding to \OI~63.1837\m\ at a redshift of $z=1.333$ which essentially matches the redshift of $z=1.334$ given by Schmidt \& Green (1983).  The estimated flux is $9.8\cdot10^{-18}~{\rm W~m}^{-2}$ at a signal-to-noise ratio of 2.7 (where the noise is computed from the line-subtracted spectrum).  This flux corresponds to $3.1\cdot10^9~L_\odot$ for a flat Universe with $H_{\rm 0}=71~{\rm km}~{\rm s}^{-1}~{\rm Mpc}^{-1}$ and $\Omega_{\rm m}=0.27$ (Spergel et al. 2003).

\section{DISCUSSION}


The only target possibly detected in this survey is the radio-quiet quasar, the other three quasars being radio-loud.  It may be notable that the marginally detected source is the only one that has an infrared-to-radio ratio that is indistinguishable from that of star-forming galaxies (Helou, Soifer, \& Rowan-Robinson 1985; Yun, Reddy, \& Condon 2001).  While this does not prove that IRAS~F16348$+$7037 is dominated energetically by star formation, the departure of the other quasars from infrared-to-radio values typical of star formation rules out star formation as the exclusive power source for them.   In light of this and the infrared-radio spectra displayed in Figure~\ref{fig:ir_radio_spectra}, one may be led to believe that perhaps there is very little interstellar dust emission in the three radio-loud sources and that their \IRAS\ flux levels merely derive from the same mechanism for the radio continuum.  The complete infrared-radio spectra, however, are not well-fitted by any combination of thermal ($f^{\rm th}_\nu \propto \nu^{-0.1}$) and non-thermal ($f^{\rm n}_\nu \propto \nu^{-0.8}$) radio continuum slopes typical of star-forming galaxies (Condon 1992).

Solomon et al. (1997) showed that local ultraluminous infrared galaxies typically are quite luminous in CO.  That CO was not detected in the three systems in this sample in which such observations were attempted (\S~\ref{sec:sample}) is consistent with the lack of a conspicuous interstellar medium.  Alternatively, a depressed metal abundance could lead to weak \OI~63\m.  But this is an unlikely explanation as quasars appear to exhibit solar metallicity at all observed redshifts (e.g. Dietrich et al. 2003; Verner et al. 2003).  On the other hand, it is possible that self-absorption is responsible for the insubstantial \OI~63\m\ emission.  Arp~220 is seen in absorption for the \OI~63\m\ line, for example (Luhman et al. 2003).  However, \OI~63\m\ emission is detected from more than half of the ultraluminous infrared galaxies observed in the sample studied by Luhman et al. (2003), and their photodissociation region modeling rules out self-absorption for the majority.  The galaxies are also likely to be heavily obscured by dust.  Luhman et al. (2003) suggest that the deficit of far-infrared line emission with respect to the continuum emission for ultraluminous infrared galaxies may be due to enhanced dust absorption in ionized gas, a circumstance that would contribute to the continuum emission but not the line emission.  Using the Li \& Draine (2001) extinction curve, $\tau_{\rm dust}(63\micron)\sim1$ corresponds to $A_V\sim300$.  Locally, a few ultraluminous infrared galaxies may exhibit this amount of extinction but most involve smaller values (Genzel et al. 1998; Genzel et al. 2001; Soifer et al. 2002; Tacconi et al. 2002).

\subsection{Comparison with Nearby Galaxies}

\begin{figure}[!ht]
\plotone{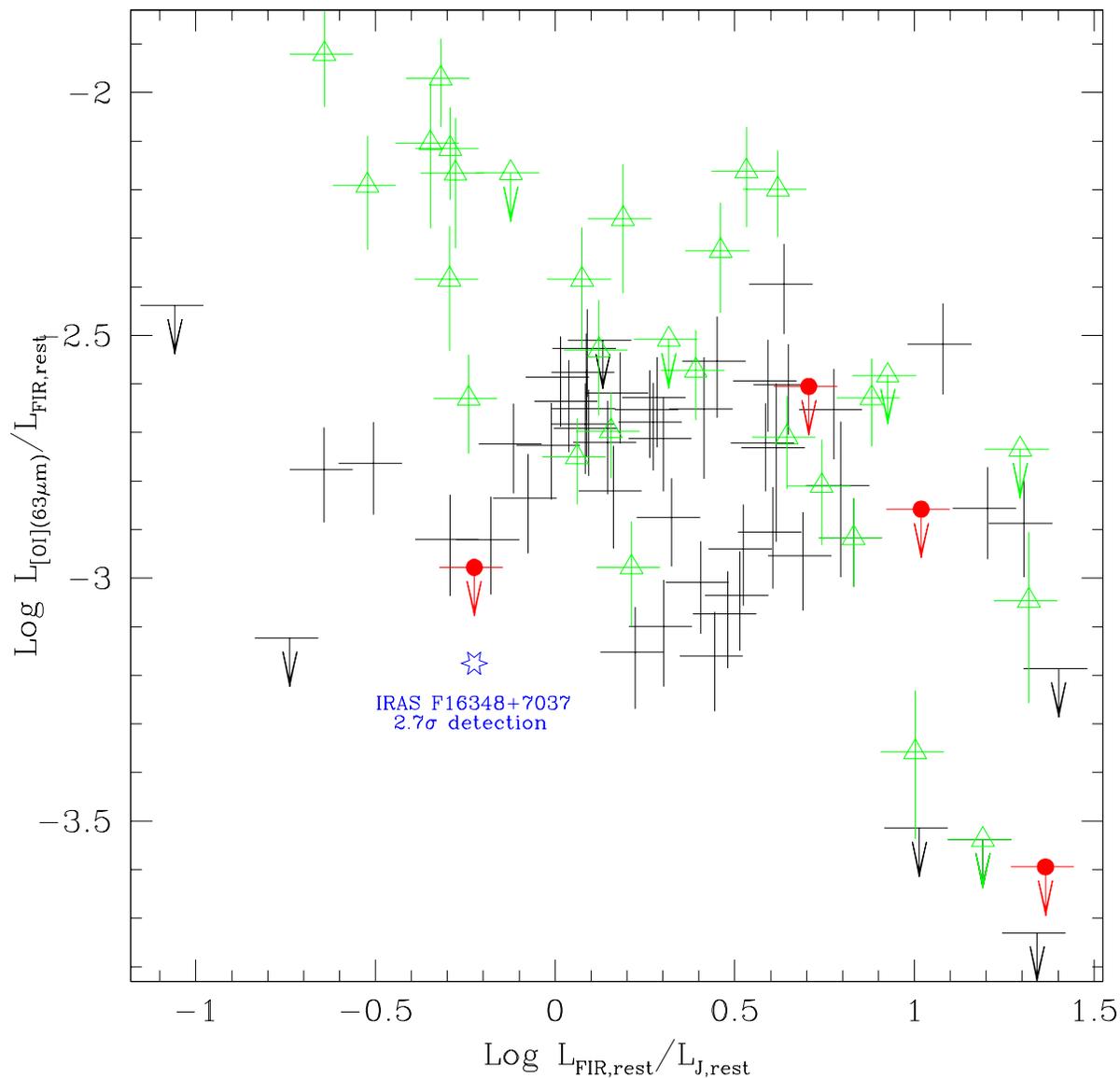}
\caption[] {\ Ratio of \OI~63\m\ luminosity to the rest-frame 42-122\m\ continuum luminosity, as a function of the rest-frame far-to-near-infrared luminosity ratio.  The data plotted with filled circles indicate the higher-redshift sources studied in this work and use 3$\sigma$ upper limits for \OI~63\m; the tentative 2.7$\sigma$ detection for IRAS~F16348+7037 is indicated by a star.  The remaining data come from the Malhotra et al. (2001) study of normal star-forming galaxies (plotted with no symbols) and Seyferts from the Brauher (2004) study (plotted with open triangles).} 
\label{fig:ratios}
\end{figure}

To put these sources in context with normal star-forming galaxies in the Local Universe, Figure~\ref{fig:ratios} displays \OI-to-$FIR$ as a function of $FIR$-to-$J$ (all estimated to rest-frame quantities) for the four sources in this program and the sample from the $ISO$ Key Project on Normal Galaxies (Malhotra et al. 2001).  Just one of the high-redshift data points lies above the mean for normal star-forming galaxies: $\left<\log~L_{[{\rm O~I}]}/L_{FIR}\right>_{\rm SF}=-$2.77$\pm$0.03 with a dispersion of 0.18~dex for the 45 normal galaxies detected in the Malhotra sample.  However, though the high-redshift data points are admittedly based only on \OI~63\m\ upper limits, the ultraluminous QSOs generally fit within the local normal galaxy envelope in this plot.  The main conclusion here is that even though these QSOs were not detected in \OI~63\m, we cannot rule out that they have substantial amounts of dense warm molecular gas undergoing  photodissociation and by extension star formation activity powering the interstellar medium and dust emission.  The data do not exclude warm molecular gas in these systems, and in the case of IRAS~F16348$+$7037, such gas may be detected.
	
An additional comparison sample was sought to help place these results in the context of active galaxies in the Local Universe.  Brauher (2004) has uniformly processed essentially all the $LWS$ data taken for extragalactic sources (except M~31 and the Magellanic Clouds).  Thirty-two of the objects studied by Brauher are listed as a Seyfert in NED and are distant enough such that the far-infrared emission fits within one $LWS$ beam ($FWHM\sim75$\arcsec).  These are mostly local Seyferts with all but one lying at a redshift smaller than 0.13; the most distant Seyfert in the sample is at $z=0.30$.  The global \OI-to-$FIR$ ratio is higher by a factor of two for these Seyferts than for the \ISO\ Key Project normal galaxies:  $\left<\log~L_{[{\rm O~I}]}/L_{FIR}\right>_{\rm Sey}=-2.46$$\pm$0.08 with a dispersion of 0.40~dex for the 25 Seyferts with \OI~63\m\ detections (the mean remains statistically the same if upper limits from the seven non-detections are included).  

\begin{figure}[!ht]
\plotone{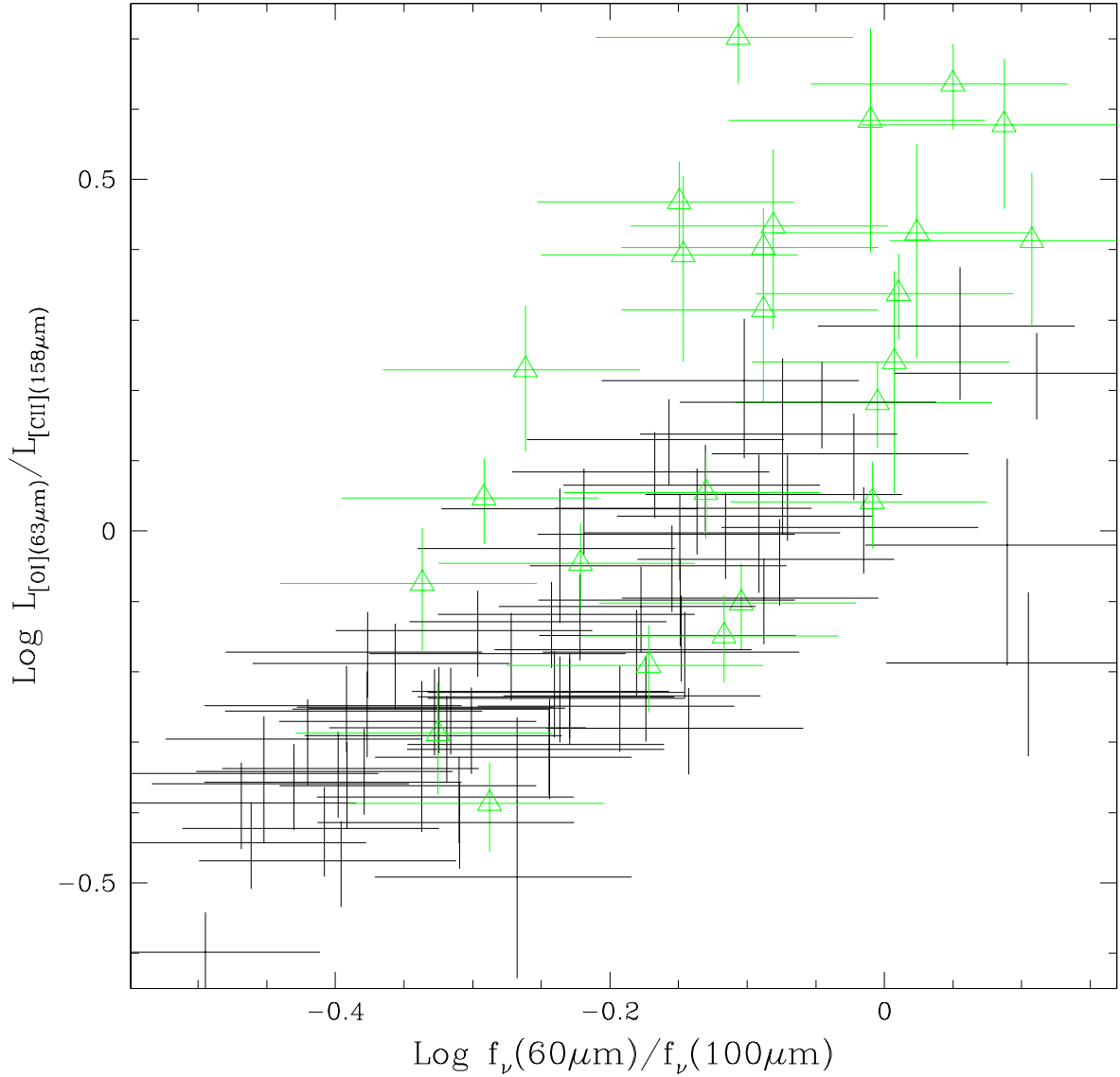}
\caption[] {\ Ratio of global \OI~63\m\ luminosity to the global \CII~158\m\ luminosity, as a function of the global \IRAS\ color \IRAScolor.  The data are from Brauher et al. (2004) with the Seyfert data plotted with open triangles.} 
\label{fig:OI_CII}
\end{figure}

The simplest approach to interpreting the data might be to associate larger relative amounts of \OI~63\m\ emission with greater amounts of mass in photodissociation regions, and by extension more abundant star formation activity.  However, the fact that Seyferts display systematically more elevated \OI/$FIR$ ratios flies directly in the face of such a simple approach.  The first empirical question to ask is whether this behavior can be understood in the context of what is known about star-forming galaxies, on the assumption that the Seyfert labels are of secondary importance.  Indeed, the Seyferts in this discussion have warmer \IRAScolor, which Malhotra et al. (2001) have shown to be associated with larger \OI/\CII\ values.  However, the Seyferts' elevated \OI\ cannot be entirely attributed to their warmer far-infrared colors:  as seen in Figure~\ref{fig:OI_CII},
the Seyferts scatter to higher \OI/\CII\ values even for warmer far-infrared colors, strongly suggesting that the line emission in the two samples may be driven by different excitation mechanisms.  For example, the Seyferts may have a stronger shock component.

The most natural explanation would be to tie the different behavior of the Seyferts to their more intense X-ray environment.  Maloney, Hollenbach, \& Tielens (1996) have modelled X-ray dissociation regions.  The emission of the interstellar medium in these regions is driven by X-ray photons from the active galactic nuclei penetrating deeply into dense circumnuclear molecular rings or giant molecular clouds in the galaxy's disk.  At higher molecular densities $n\sim10^5~{\rm cm}^{-3}$ and higher illuminations \OI~63\m\ emission can exceed \CII~158\m\ by a substantial margin, and carry a large fraction of the total far-infrared dust emission.  Maloney, Hollenbach, \& Tielens (1996) predict a
\OI\ line-to-continuum ratio in X-ray dissociation regions that is substantially larger than for the photodissociation regions that comprise the bulk of the interstellar mass in normal galaxies.

\subsection{Implications for Future Work}


The results of this work can help to better develop future surveys for far-infrared emission lines.  As outlined in \S~1, the \CII~158\m\ line should dominate the cooling of the interstellar gas for relatively cool galaxies whereas \OI~63\m\ becomes increasingly more important for warmer galaxies.  Cold, ultraluminous objects do exist at high redshifts (e.g. Chapman et al. 2002) and will be favored in future \CII~158\m\ surveys of distant galaxies, e.g. surveys carried out at $z\sim3-5$ with the Atacama Large Millimeter Array.  Though \OI~63\m\ has been detected in this survey from at most one of four ultraluminous galaxies, local Seyferts show relatively elevated \OI~63\m\ emission.  Future high-redshift \CII~158\m\ surveys should be complemented by \OI~63\m\ observations.  Combining \CII\ and \OI\ observations will reveal a fuller picture of early nucleosynthesis and interstellar processes in star-forming galaxies, and will highlight X-ray-driven processes in active galaxies.  Efforts to detect \OI~63\m\ will require high resolution spectrometers sensitive to 240\m~$\lesssim \lambda \lesssim$ 400\m\ for $3 \lesssim z \lesssim 5$, capabilities planned for the {\it Herschel Space Observatory}.  


\section{SUMMARY}

A search has been carried out for a marker of warm and dense interstellar gas, the \OI~63\m\ emission line, in four ultraluminous QSOs at redshifts between 0.6 and 1.4.  The observations achieved a sensitivity sufficient to detect line emission in comparably-luminous galaxies dominated by star formation.  No robust detections are reported, but one source, IRAS~F16348$+$7037, may exhibit emission at a signal-to-noise ratio of 2.7.  In terms of the parameters \OI/$FIR$ and $FIR$/$L_J$, the observed upper limits place the quasars within the general envelope that local galaxies define.  It is likely that the tentative detection for IRAS~F16348$+$7037, coupled with its conspicuous far-infrared peak in its long-wavelength spectrum, imply that this particular high-luminosity, radio-quiet quasar has a substantial interstellar medium.  With the data obtained for the three radio-loud quasars we can only suggest that the presence of warm molecular gas cannot be ruled out for these systems.  Comparisons with far-infrared line emission from local galaxies show that Seyfert galaxies exhibit elevated \OI~63\m\ emission, a finding that may be explained by the enhanced X-ray emission from active nuclei.  Regardless of any possible evolutionary links between Seyferts, ultraluminous, and normal galaxies, the results presented here point to \OI~63\m\ as an important cooling line for a wide range of interstellar environments.

\acknowledgements 
We would like to thank M. Brotherton for useful comments.  This publication makes use of data products from the Two Micron All Sky Survey, which is a joint project of the University of Massachusetts and the Infrared Processing and Analysis Center/California Institute of Technology, funded by the National Aeronautics and Space Administration and the National Science Foundation.  This research has made use of the NASA/IPAC Extragalactic Database which is operated by JPL/Caltech, under contract with NASA.  This research has made use of the NASA/IPAC Extragalactic Database which is operated by JPL/Caltech, under contract with NASA.  The Digitized Sky Surveys were produced at the Space Telescope Institute under U.S. Government grant NAG~W-2166.  The images of these surveys are based on photographic data obtained using the Oschin Schmidt Telescope on Palomar Mountain and the UK Schmidt Telescope.

\clearpage

\scriptsize
\begin{deluxetable}{cccccccc}
\def\a{\tablenotemark{a}}

\def\p{$\pm$}
\tablenum{1}
\label{tab:sample}
\tablewidth{500pt}
\tablecaption{The Sample\a}
\tablehead{
\colhead{Galaxy} &
\colhead{Type} &
\colhead{R.A.} &
\colhead{Decl.} &
\colhead{$z$} &
\colhead{$f_\nu(1.4$GHz)} & 
\colhead{$\log {L_{FIR} \over L_\odot}$} & 
\colhead{$\log {L_{FIR} \over L_J}$} 
\\
\colhead{} &
\colhead{} &
\colhead{(J2000)} &
\colhead{(J2000)} &
\colhead{} &
\colhead{(mJy)} & 
\colhead{} & 
\colhead{} 
}
\startdata
IRAS~F04207$-$0127&QSO blazar   &042315.8&$-$012033&0.914&2700\p~80&13.4&$+$1.02\\
IRAS~F16348$+$7037&QSO~~~~~~~~~~&163428.9&$+$703133&1.334&0.94\p0.3&13.6&$-$0.22\\
IRAS~F16413$+$3954&QSO blazar   &164258.8&$+$394837&0.593&7100\p200&13.2&$+$0.71\\
IRAS~F22231$-$0512&QSO blazar   &222547.3&$-$045701&1.404&7400\p300&14.4&$+$1.37\\
\enddata
\tablenotetext{a}{\footnotesize Type, position, redshift, and radio flux density are drawn from NED.  The estimation of rest-frame quantities $L_{FIR}$ and $L_J$ are described in the text.}
\end{deluxetable}
\normalsize

\scriptsize
\begin{deluxetable}{ccccccc}

\tablenum{2}
\label{tab:observations}
\tablewidth{480pt}
\tablecaption{\ISO\ \OI~63\m\ Observations}
\tablehead{
\colhead{Galaxy} &
\colhead{Observation ID} &
\colhead{$\lambda_{\rm obs}$} &
\colhead{$T_{\rm On}$} &
\colhead{$T^{\rm a}_{\rm Off}$} &
\colhead{continuum rms} &
\colhead{$3\sigma~f($\OI)} 
\\
\colhead{} &
\colhead{($ISO$ TDT)} &
\colhead{(\micron)} &
\colhead{(sec)} &
\colhead{(sec)} &
\colhead{{\footnotesize($10^{-17}~$W/m$^2/\micron$)}} &
\colhead{{\footnotesize($10^{-17}~$W/m$^2$)}} 
}
\startdata
IRAS~F04207$-$0127&66402450-2~~~~~~&120.9&19820&1398&1.81&$<$~3.5\\
IRAS~F16348$+$7037&52100220-2,4-6  &147.5&13264&1180&0.80&$<$~1.9\\
IRAS~F16413$+$3954&59501840-3~~~~~ &100.6&~7367&1178&5.59&$<$10.~~\\
IRAS~F22231$-$0512&53502130-4~~~~~ &151.9&~8770&1176&1.02&$<$~2.5\\
\enddata
\tablenotetext{a}{\footnotesize Not used in computing continuum r.m.s.}
\end{deluxetable}

\clearpage

\normalsize

\end{document}